\documentclass[twocolumn,preprintnumbers,amsmath,amssymb]{revtex4-1}
\usepackage{graphicx}
\usepackage{ulem}
\usepackage[svgnames]{xcolor}
\usepackage{bm}
\usepackage{multirow}
\usepackage{amsmath}
\newif\ifshowcomments\showcommentstrue

\begin{document}

\title{The free energy of twisting spins in Mn$_3$Sn}

\author{Xiaokang Li$^{1,*}$, Shan Jiang$^{3,1}$, Qingkai Meng$^{1}$, Huakun Zuo$^{1}$, Zengwei Zhu$^{1,*}$, Leon Balents$^{2,4}$ and Kamran Behnia$^{3}$} 

\affiliation{(1) Wuhan National High Magnetic Field Center and School of Physics, Huazhong University of Science and Technology, Wuhan 430074, China\\
(2)Kavli Institute for Theoretical Physics, University of California, Santa Barbara, California 93106-4030, USA \\ 
(3) Laboratoire de Physique et d'\'Etude des Mat\'eriaux \\ (ESPCI - CNRS - Sorbonne Universit\'e), PSL Research University, 75005 Paris, France\\ (4) Canadian Institute for Advanced Research, Toronto, Ontario, Canada}

\date{\today}
\begin{abstract}

The magnetic free energy is usually quadratic in magnetic field and depends on the mutual orientation of the magnetic field and the crystalline axes. Tiny in magnitude, this magnetocrystalline anisotropy energy (MAE) is nevertheless indispensable for the existence of permanent magnets. Here, we show that in Mn$_3$Sn, a non-collinear antiferromagnet attracting much attention following the discovery of its large anomalous Hall effect, the free energy of spins has superquadratic components, which drive the MAE. We experimentally demonstrate that the thermodynamic free energy includes terms odd in magnetic field ($\mathcal{O}(H^3)+\mathcal{O}(H^5)$) and generating sixfold and twelve-fold angular oscillations in the torque response. We show that they are quantitatively explained by theory, which can be used to quantify relevant energy scales (Heisenberg, Dzyaloshinskii-Moriya, Zeeman and single-ion anisotropy) of the system. Based on the theory, we conclude that, in contrast with common magnets, what  drives the MAE in Mn$_3$Sn is the field-induced deformation of the spin texture.
\end{abstract}
\maketitle
Aligned spins located on two adjacent atoms are affected by the anisotropic electrostatic forces connecting their orbital angular momenta~\cite{vanVleck}. This magnetocrystalline anisotropy energy (MAE), a consequence of the spin-orbit coupling, is remarkably small ($\sim 60 \mu$ eV/atom in Co and $\sim 1 \mu$ eV/atom in Fe and Ni). Since it is the outcome of the competition between energies many orders of magnitude larger, it is hard to calculate from first principles~\cite{Daalderop1990,Halilov1998}.

 Mn$_3$Sn, a noncollinear antiferromagnet with an inverse triangle spin structure located on a breathing kagome lattice~\cite{Tomiyoshi1982} has attracted much attention following the observation of a large anomalous Hall effect(AHE) ~\cite{Nakatsuji2015} with a sizeable net Berry curvature near the Fermi level~\cite{Yang2017}. The discovery was followed by the observation of various counterparts of AHE, including the anomalous Nernst~\cite{Ikhlas2017, Li2017} and the anomalous thermal Hall effects~\cite{Li2017, Sugii2019, Xu2020}, as well as the anomalous magneto-optical Kerr effect~\cite{Higo2018, Balk2019}. These are room-temperature effects requiring a small magnetic field. Therefore, Mn$_3$Sn is potentially attractive in the field of antiferromagnetic spintronics~\cite{Smejkal2018, Baltz2018, Kimata2019, Tsai2020} or as a Nernst thermopile~\cite{Ikhlas2017, Li2021}. The peculiar spin texture of Mn$_3$Sn has been subject of several studies~\cite{Cable1993, Liu2017, Suzuki2017, Park2018, Zelenskiy2021, Miwa2021}. The magnetic Hamiltonian includes Heisenberg and Dzyloshinskii-Moriya spin-spin interaction terms dominating by far the small single-ion anisotropy term~\cite{Liu2017}. A study of torque magnetometry~\cite{Duan2015} quantified the latter. Previous experiments have documented that magnetic domain walls are chiral~\cite{Li2018} and host a topological Hall effect associated with a finite skyrmionic number~\cite{Li2017}. 
 
Here, we present a study of magnetization and angle-dependent magnetic torque in Mn$_3$Sn single crystals and find an additional term in the magnetic free energy  with an intriguingly delicate anisotropy. Magnetization, on top of its zero-field (spontaneous) and field-linear components, has a previously undetected term, which is quadratic in magnetic field. This implies that the free energy has a component proportional to the cube of the magnetic field~\cite{Treves1962}.  We show that, in contrast to ordinary ferromagnets, this second-order magnetization is the driver of the magneto-crystalline anisotropy. This conclusion is based on our torque magnetometry data, which uncovers a field-dependent MAE. The angular oscillations of the torque signal in addition to its two-fold, $K_2$, and sixfold, $K_6$, components detected previously~\cite{Duan2015}, have an additional twelve-fold, $K_{12}$, component. The field dependence of the two intrinsic shape independent terms ($K_6 \propto H^3$ and $K_{12} \propto H^5$ ) establish that they are caused by the second-order magnetization.   Both the six-fold and twelve-fold anisotropies are quantitatively reproduced by a theoretical calculation based on the Hamiltonian of ref. \cite{Liu2017}. Fits to the experimental quantifies the Zeeman energy, the Heisenberg (J) and  the Dzyaloshinskii-Moriya (DM) interaction term, and finally the single-ion-anisotropy energy, the cost of misalignment between spin lattice and crystal axes. The analysis demonstrates that the unusual features are driven by field-induced twisting of the spins, which result from the frustrated competition of the above four energy scales, particular to the inverse triangular spin structure.

\begin{figure*}
\centering
\includegraphics[width=16cm]{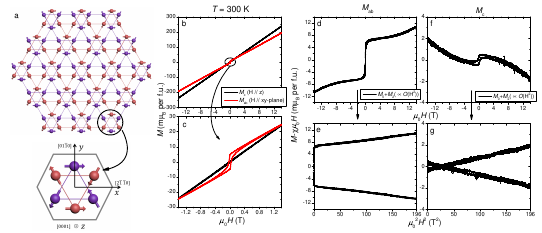}
\caption{\textbf{A quadratic term in the magnetization of Mn$_3$Sn:} (a) The spin texture of Mn$_3$Sn. Mn magnetic moments are locating on an AB-stacked breathing kagome lattice. (b-c) In-plane, $M_{ab}$, and the out-of-plane, $M_{c}$, magnetization in Mn$_3$Sn at 300 K, as a function of magnetic field swept from 14 T to -14 T. (d-e) The residual component of $M_{ab}$ after subtracting its dominant linear term $M_1(\propto O(H))$. It is plotted with the linear (d) and quadratic (e) $x$-coordinate respectively. There is an additional component,  $M_2(\propto O(H^2))$. (f-g) The residual component of $M_{c}$. The sign of its second-order magnetization is opposite to $M_{ab}$.}
\label{fig: Magnetization}
\end{figure*}

\begin{figure}
\centering
\includegraphics[width=8cm]{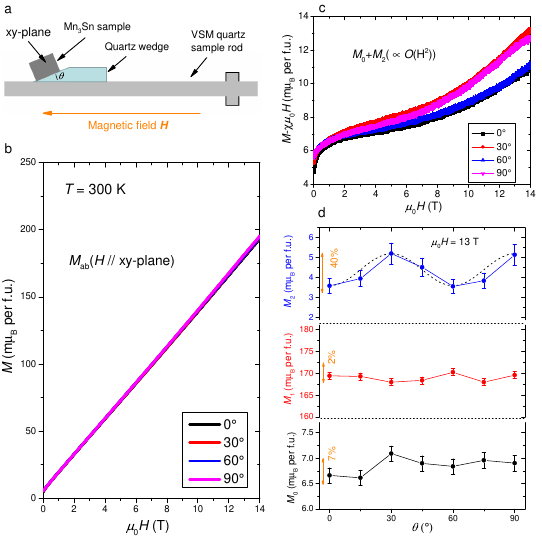}
    \caption{\textbf{Anisotropy of the quadratic magnetization :} (a) Experimental configuration for the angular magnetization. (b-c) Comparison of $M_{ab}$ and it's residual after subtracting the linear background, with four different angles, 0$^\circ$, 30$^\circ$, 60$^\circ$ and 90$^\circ$ respectively. (d) The angle dependence of the $M_0$, $M_1$, and $M_2$. Only the latter show angular oscillations with a periodicity of 60 degrees.}
\label{fig: Angular quadratic magnetization}
\end{figure}

Centimeter-size Mn$_3$Sn single crystals were grown by the vertical Bridgman technique~\cite{Li2018}. As shown in Fig.~\ref{fig: Magnetization}a,  Mn atoms form a breathing kagome lattice. In this inverse triangle spin lattice~\cite{Tomiyoshi1982, Nakatsuji2015}, large magnetic moments of the Mn atoms cancel each other. As seen in Fig.~\ref{fig: Magnetization}b-c, a weak, and mainly in-plane, ferromagnetism persists, which may be caused by a slight deviation of spin orientations from a perfect  triangular alignment~\cite{Tomiyoshi1982, Nakatsuji2015}. A collection of six Mn atoms located in two adjacent layers constitutes the simplest unit of the spin texture and has been treated as a magnetic octupole spin cluster~\cite{Suzuki2017}.

Our first new finding is shown in Fig.~\ref{fig: Magnetization}d-g. After subtracting a linear background ($M_1(\propto\mathcal{O}(B)$), we find that in addition to the spontaneous magnetization, M$_0$, there is an additional field-dependent term (Fig.~\ref{fig: Magnetization}d) in magnetization. As seen in Fig.~\ref{fig: Magnetization}e, this term is quadratic in magnetic field. Such a term is also present in the out-of-plane magnetization too, albeit with a smaller amplitude and an opposite sign (See Fig.~\ref{fig: Magnetization}f-g). We conclude that the magnetization consists of three terms:

\begin{equation}\label{total-magnetization}
M_{total} = M_0 + M_1(\propto\mathcal{O}(H)) + M_2 (\propto \mathcal{O}(H^2)).
\end{equation}
The first term is the zero-field spontaneous magnetization associated with the weak residual ferromagnetism. The second term  ($M_1$) is the dominant linear magnetization resolved in previous studies of magnetization~\cite{Nakatsuji2015}. The third term, $M_2$, not detected in previous studies is quadratic in field. It represents a second-order correction to the magnetization response~\cite{Treves1962,Gorodetsky1964, Kharchenko1995, Liang2017}. Since the magnetization  is the partial derivative of the magnetic free energy with respect to the magnetic field ($M= \partial F_M / \partial H$), a finite $M_2$ implies an additional term for the magnetic free energy, which can be written as:

\begin{equation}\label{magnetic-free-energy}
F_M = \sum_{i} M_{0,i}H_i + 
\frac{1}{2} \sum_{i,j} \chi_{i,j} H_{i} H_{j} + 
\frac{1}{3} \sum_{i,j,k} C_{i,j,k} H_{i} H_{j} H_{k}.
\end{equation}

Here, $C_{i,j,k}$ is a $3\times3\times3$ tensor, which represents the second odd term in the field dependence of the free energy. 

The angular variation of these three terms was investigated by performing the angular dependent magnetization measurements, using the set-up shown in Fig.~\ref{fig: Angular quadratic magnetization}. Rotation was achieved  by changing the sharp angle of a quartz wedge held between the sample and the sample-holder. Fig.~\ref{fig: Angular quadratic magnetization}b and c show magnetization and its residual after subtracting the linear background as a function of field for four different angles in $xy$-plane. 

The angle dependence of in-plane $M_0$,  $M_1$, and  $M_2$  are plotted in Fig.~\ref{fig: Angular quadratic magnetization}d. Angular variation is undetectable for $M_0$ and $M_1$, but as large as 37 percent for $M_2$ and displays  a clear six-fold symmetry, with minima and maxima separated by 30 degrees. We conclude that the third term of free energy $F_3\propto O(H^3)$ has a sixfold anisotropic component. In other words, $M_2= M_{2,0}+  M_{2,6} cos(6\phi)$. At 13 T, $M_{2,0}= 4.4 ~m\mu_B$/f.u., and $M_{2,6}= 0.8 ~m\mu_B$/f.u. As we will see below, the magnitude of the latter term is confirmed with a higher precision by measurements of  the magnetic torque, $\tau$, which quantifies the angular derivative of the magnetic free energy($\tau= \partial F_M / \partial \theta$).
\begin{figure}
\includegraphics[width=8cm]{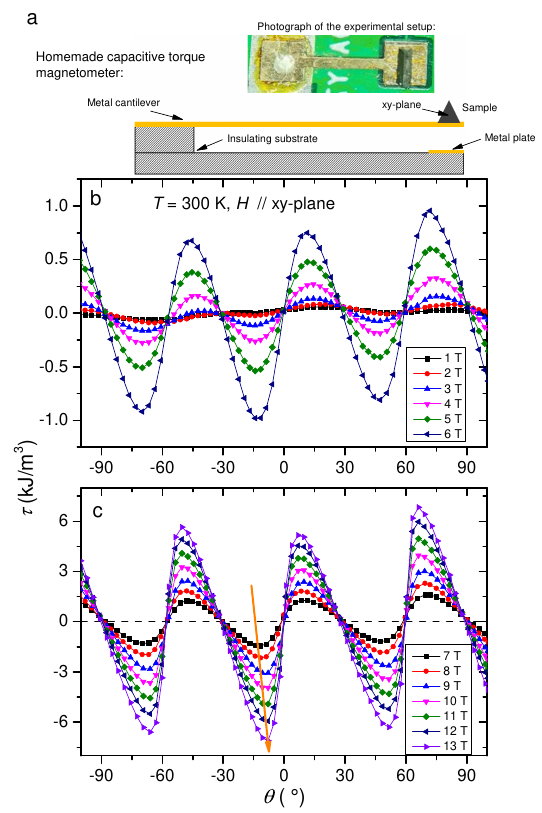}
    \caption{\textbf{Magnetic torque measurements:} (a) The home-made experimental setup and its photograph (top) with a capacitive torque magnetometer, rotating the field in the $xy$-plane of the Mn$_3$Sn sample. (b) The angle dependent torque responses for magnetic fields up to 6 T. (c) The angle dependent torque responses for magnetic fields larger than 6 T. As the field increases, the shape of oscillations evolve, indicating the emergence of an additional component. $H \parallel x$ corresponds to 0$^\circ$.}
\label{fig: Torque magnetometer}
\end{figure}

\begin{figure*}
\includegraphics[width=16cm]{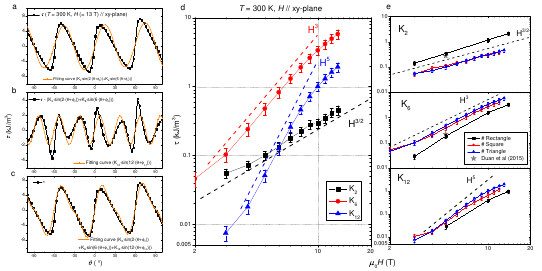}
    \caption{\textbf{Components of the torque response:} 
    (a) Fit to the 13 T data with an expression, which has only a two-fold and a six-fold term ($\tau = K_2 \cdot sin(2 \cdot (\phi + \phi_2)) + K_6 \cdot sin(6 \cdot (\phi + \phi_6))$). The mismatch is obvious. (b) The residual torque component after subtracting the data and the previous fit. It shows a clear twelve-fold symmetry. (c) A fit which includes an additional twelve-fold symmetry component ($K_{12} \cdot sin(12 \cdot (\phi + \phi_{12})$)). (d) Fitting parameters $K_2(H)$, $K_6(H)$ and $K_{12}(H)$ as a function of magnetic field.  $K_2(H)$ follows $H^{3/2}$ and is much smaller than $K_6(H)$. $K_6(H)$ follows $H^{3}$. $K_{12}(H)$ follows $H^{5}$ at the beginning. (e) A comparison of $K_2(H)$, $K_6(H)$ and $K_{12}(H)$ in three different samples with different cross sections. Here, the  sample with the rectangular cross section has a larger aspect ratio ($l_x/l_y$ = 5.6), compared to the samples with square and triangle cross sections ($l_x/l_y \simeq 1$). The data  reported by Duan \textit{et al.} are also shown.}
\label{fig: Magnetic anisotropic energy}
\end{figure*}

\begin{figure*}
\centering
\includegraphics[width=16cm]{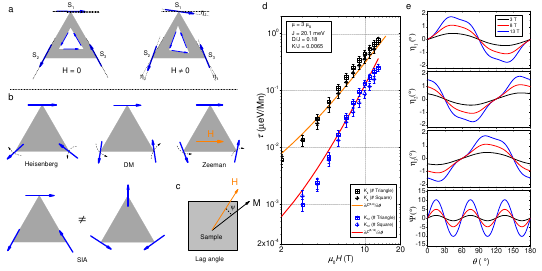}
   \caption{\textbf{Magneto-crystalline anisotropy driven by field-induced twist of non-aligned spins:} (a) In absence of magnetic field, the spin lattice is triangular. The magnetic field distorts the spin triangle (in white), which is no more isomorphic to the lattice triangle (in gray). The deformation angles $\eta_{i}$ quantify the distortion. (b) Interaction between one spin and its immediate neighbors favor clockwise and anticlockwise twists. The Zeeman effect favors alignment of all spins with magnetic field. Single-ion anisotropy causes in-equivalency between the two perpendicular orientations of the spin triangle with respect to the lattice triangle.  (c) When the magnetic field rotates, there is a lag angle $\psi$ between the  magnetic field and the total magnetization . (d) The experimental $K_6$ and $K_{12}$ (symbols) compared to theoretical expectation (solid line) using $ \mu = 3 \mu_B$, $J = 20.1 meV$, $D/J = 0.18$, $K/J = 0.0065$. (e) The deformation angles $\eta_{i}$ and lag angle $\psi$ at 3 T, 8 T and 13 T predicted by theory \cite{SM}. The three tilt angles follow the same pattern with  an angular delay of $60^{\circ}$. With increasing magnetic field, a secondary oscillation with a periodicity of $30^{\circ}$ emerges on top of a $60^{\circ}$ periodicity.}
\label{fig: Mechanism}
\end{figure*}

Fig.~\ref{fig: Torque magnetometer}a shows a sketch and a photograph of the torque set-up. The applied magnetic field rotates in the ab-plane of the Mn$_3$Sn sample, and the magnetic torque along the c-axis is detected by the variation in the capacitance between two metal plates. Fig.~\ref{fig: Torque magnetometer}b shows the angular dependence of the torque signal below 6 T resolving a clear six-fold symmetry. However, as the magnetic field increases further, a deviation becomes visible. As seen in Fig.~\ref{fig: Torque magnetometer}c, when the magnetic field increases from 6 T to 13 T, the shape of each oscillation evolves. Instead of being sine-like, it becomes sawtooth-like. When the field is along a high symmetry axis ($x,-x,y,-y$), torque vanishes and the free energy is at its extremum, as one expects. Remarkably, however, there is a widening difference between the slopes along the two high-symmetry orientations. This is caused by the emergence of an additional angle-dependent term. 

In addition to the two-fold and six-fold terms detected by Duan \textit{et al.}\cite{Duan2015}, our high-field data has an additional term with twelve-fold symmetry. This can be seen in Fig.~\ref{fig: Magnetic anisotropic energy}a, which shows that the difference between the experimental data and the sum of the two first terms ($K_2 \cdot sin(2 \cdot (\theta + \phi_2)) + K_6 \cdot sin(6 \cdot (\theta + \phi_6))$). The residual has a clear twelve-fold anisotropy ($K_{12} \cdot sin(12 \cdot (\theta + \phi_{12}))$). The evolution of the fitting parameters, $K_2$, $K_6$, and $K_{12}$  with magnetic field is shown in Fig.~\ref{fig: Magnetic anisotropic energy}d and e. $K_2(H)$ follows $H^{3/2}$ and $K_6(H)$ follows $H^{3}$ and remains the largest in the whole field range. Finally, $K_{12}(H)$ shows a steep increase, which is close to $H^{5}$ at low fields.

The strong field dependence of the torque components of Mn$_3$Sn is in sharp contrast with what is seen in ferromagnets\cite{Graham1958}, as confirmed by our own data on bcc Fe (See the supplement\cite{SM}). Additionally, it implies that a single magnetic field measurement (5 T in the case of ref.~\cite{Duan2015}, see Fig.\ref{fig: Magnetic anisotropic energy}e) is not sufficient to extract the magnotcrystalline anisotropy.  Let us now consider the information brought by these results on the components of magnetic free energy.

The two-fold term, $K_2$, evolves much slower than $H^3$ and strongly depends on the sample aspect ratio (See Fig.~\ref{fig: Magnetic anisotropic energy}e). The first feature indicates that it is not caused by the second-order magnetization and the second feature implies a role played by sample geometry. $K_2$ can be caused by a finite angle between the applied field and magnetization, which can, in principle, occur in any sample with finite size. 

The six-fold term, $K_6$, shows a field dependence clearly linking it to the field-cubic free energy ($F_3\propto\mathcal{O}(H^3)$), confirming what was deduced from measurements of angle-dependent magnetization. Both sets of data lead to the conclusion that there is a component of free energy with sixfold angular symmetry and cubic field dependence ($F^{(3,6)}\propto H^3cos6\theta$). A consistency check can be done by comparing the magnitude of $K_6$, the angle derivative  and $M_{2,6}$, the field derivative. At 13 T, $K_6$ is  4700- 5600 $Jm^{-3}$,  implying $F^{(3,6)}\approx 0.34\pm4 ~\mu eV/f.u.$ and $M_{2,6}$ is $ 0.8 ~m\mu_B$ per f.u. corresponding to $F^{(3,6)}\approx 0.20 ~\mu eV/f.u. $.  The list of all components of the free energy identified by our experiments are given in Table~ \ref{Table-summary}.

 \begin{table*}
 \centering
 \begin{ruledtabular}
\begin{tabular}{c|c|c|c|c|c|}
 & Component  & Gives rise to & Amplitude at 13 T ($\mu$ eV/Mn) & Accounted by theory? & Comments  \\
 \hline
& $F^{(1,ab)}\propto H$ & M$_0$ & 1.7  & Yes & residual FM  \\   
\hline
& $F^{(1,c)}\propto H$ & M$_{0}$ & 0.11  & Canting? & residual FM   \\   
\hline
& $F^{(2,ab)}\propto H^2$ & M$_1$ & 21.4   & Yes & paramagnetism \\ 
\hline
& $F^{(2,c)}\propto H^2$ & M$_1$ & 27.9   & Yes & paramagnetism \\ 
\hline
& $F^{(3,0)}\propto H^3$ & M$_{2,0}$ & 0.37 & No & Field-induced correction to FM?\\ 
\hline
& $F^{(3,6)}\propto H^3cos6\theta$ & M$_{2,6}/K_{6}$& 0.067/0.115 & Yes & Field-induced tilt of spins\\ 
\hline
& $F^{(5,12)}\propto H^5cos12\theta$ & $K_{12}$& 0.018 & Yes & secondary correction to the tilt\\
\hline
& $F^{(1.5,2)}\propto H^{3/2}cos2\theta$ & $K_{2}$ & 0.028 & No & Boundary-related \\
\hline
\end{tabular} 
\end{ruledtabular}
 \caption{Components of the magnetic free energy in Mn$_3$Sn identified by measurements of magnetization ($M_{0}$, $M_{1}$ and $M_{2}$) and torque ($K_{2}$, $K_{6}$ and $K_{12}$). }
\label{Table-summary}
\end{table*}

Let us now show that theory provides a satisfactory account of the existence and the amplitude of $K_6$ term as well as the emergence and rapid growth of the secondary $K_{12}$ term with increasing magnetic field. 

Following  Liu and Balents~\cite{Liu2017}, the energy  per magnetic unit cell (six spins)  consists of the sum of four terms\cite{SM}:  These are Heisenberg: $ 4J\sum_{i}\bm{S}_i\cdot\bm{S}_{i+1}$; Dzyaloshinskii-Moriya (DM): $4D \sum_{i}\bm{\hat{z}}\cdot \bm{S}_i \times \bm{S}_{i+1}$; Single-ion-anisotropy (SIA) : $-2K \sum_{i}(\bm{S}_i\cdot\bm{\hat{e}}_i)^2 $; and Zeeman: $-2\mu \sum_{i}\bm{H}\cdot \bm{S}_i$. For $D>0$ and in absence of SIA and Zeeman terms, the ground state is an anti-chiral state with in-plane spins. A finite magnetic field  will distort the spin triangles. The distortion angles, $\eta_i$  (See Fig.\ref{fig: Mechanism} ) are small, because in our window of investigation ($H< 14$T), one has $K \ll J$ and $\mu H \ll J$. Introducing a small parameter $r\ll 1$, with $K/J, ~\mu H/J$ of $O(r)$, one can expand $\eta_{1,2}$ in a formal series in $r$ :
\begin{equation}
  \label{eq:Exp}
  \eta_i = \sum_{n=1}^\infty \eta_{i,n} r^n,
\end{equation}
By minimizing the energy terms over $\eta_{i,n}$, the small distortions of the triangle at each order can be quantified. This leads to an expression for the free energy per unit cell. The first term is linear in magnetic field (See the supplement\cite{SM}):
\begin{equation}
\label{eq:F1}
F^{(1,ab)}=  \frac{K\mu H}{J +\sqrt{3}D}
\end{equation}
The quadratic term \cite{SM} has slightly different expressions for in-plane and out-of-plane orientations of magnetic field is:
\begin{equation}
\label{eq:F2}
\begin{split}
F^{(2,ab)}=  \frac{(\mu H)^2}{2J}(1 -\frac{\sqrt{3}D}{J}) \\
F^{(2,c)}=  \frac{(\mu H)^2}{2J}(1 -\frac{D}{\sqrt{3}J}) 
\end{split}
\end{equation}

Therefore, one expects the quadratic free energy  to be larger for the out-of-plane orientation of the magnetic field, in agreement with what is seen experimentally (See Table~\ref{Table-summary}). For in-plane configuration, the first correction to the quadratic term has a $cos6\theta$ angle dependence. Its amplitude is equal to:
\begin{equation}
\label{eq:F3} 
F^{(3,6)}=   \frac{(K+\mu H)^2((3J+7\sqrt{3} D)K+4\sqrt{3} D\mu H)}{36(J+\sqrt{3}D)^3} 
\end{equation}

The field dependence of this term is close to $H^3$. As one can see in Fig. \ref{fig: Mechanism}d, this expression provides an excellent account of the field and angular dependence of the experimentally observed $K_6$. The next term has a $sin^2 (6\theta)$ angle dependence whose amplitude is equal to :
\begin{equation}
\label{eq:F5} 
\begin{split}
F^{(5,12)}= \frac{(K+\mu H)^2}{72(J+\sqrt{3}D)^5\mu HK}
((3J+7\sqrt{3}D)K^2+\\
2(J+4\sqrt{3}D)\mu HK+2\sqrt{3} D(\mu H)^2)^2
\end{split}
\end{equation}
The field dependence of this term is close to $H^5$ and it accounts for the emergence of $K_{12}$ in the torque data and its field dependence. (See Fig. \ref{fig: Mechanism}d).

The agreement between theory and experiment allows us to extract the energy scales of the system. Taking the magnetic moment of each Mn atom to be $\mu = 3 ~\mu_B$, as reported by neutron diffraction studies \cite{Brown1990, Tomiyoshi1982b}, we extracted $J$, $D$, and $K$.  The results are summarized in table~\ref{table-scales}. It shows what is yielded by fitting the torque data with equations \ref{eq:F3} and \ref{eq:F5}. Alternatively, one can use the magnetization data and equations \ref{eq:F1} and \ref{eq:F2}, the results are given in the second row of table~\ref{table-scales}.  As seen in the table~\ref{table-scales}, $J= J_1 + J_2 = 20.1$ meV, which is a third larger than what is yielded by magnetization. The most plausible source of this difference is the existence of a finite orbital contribution to magnetization \cite{Sandratskii1996} and assuming a 30 percent orbital contribution to M$_1$ would lead to consistency. We note that our result is fairly close to the what has been reported by a study of magnon dispersion by inelastic neutron scattering (18 meV)\cite{Park2018}. Finally, our study pins down the values for $K$ and $D$ and demonstrates the capacity of torque magnetometry to quantify the magnitude of Dzyaloshinskii-Moriya interaction in a frustrated magnet.

\begin{table}[htb]
\begin{tabular}{|c|c|c|c|c|}
\hline
Parameter & J (meV) & D/J & K/J  \\
\hline 
Torque & 20.1(8) & 0.18(2) & 0.0065(5) \\
\hline
Magnetization & 13.8 & 0.18 & 0.0058 \\
\hline
\end{tabular}
\caption{The energy scales of Mn$_3$Sn extracted from the torque and magnetization data using the theoretical expressions of the free energy.}
\label{table-scales}
\end{table}

In summary, we resolved different components of the magnetic free energy in Mn$_3$Sn. In addition to the dominant term, which is even in magnetic field, it includes odd terms with superquadratic field dependence and presenting sixfold and twelvefold angular oscillations. We showed that theory invoking a field-induced twist in the orientation of in-plane spins can successfully explain the presence of these terms and their amplitude can be used to extract all energy scales of the system.  The model also makes experimental predictions of specific twists which may be tested in the future in neutron scattering and anomalous Hall effect measurements.  It is remarkable that the simple localized spin description used here works in such quantitative detail for this highly itinerant magnet.  

This work was supported by the National Science Foundation of China (Grant No.51861135104 and No.11574097) and The National Key Research and Development Program of China (Grant No.2016YFA0401704). K. B was supported by the Agence Nationale de la Recherche (ANR-18-CE92-0020-01; ANR-19-CE30-0014-04). S. J. acknowledges a PhD scholarship by the China Scholarship Council (CSC). X. L. acknowledges the China National Postdoctoral Program for Innovative Talents (Grant No.BX20200143) and the China Postdoctoral Science Foundation (Grant No.2020M682386).  L.B. was supported by the NSF CMMT program under Grant No. DMR-2116515.

\noindent
* \verb|lixiaokang@hust.edu.cn|\\
* \verb|zengwei.zhu@hust.edu.cn|\\

\bibliography{main}

\clearpage
\onecolumngrid
\renewcommand{\thesection}{S\arabic{section}}
\renewcommand{\thetable}{S\arabic{table}}
\renewcommand{\thefigure}{S\arabic{figure}}
\renewcommand{\theequation}{S\arabic{equation}}
\setcounter{section}{0}
\setcounter{figure}{0}
\setcounter{table}{0}
\setcounter{equation}{0}
\begin{center}{\large\bf Supplemental Material for ``The free energy of twisting spins in Mn$_3$Sn"}\\
\end{center}
\setcounter{figure}{0}


\section{Samples and Methods} Mn$_3$Sn single crystals used in this work were cut to desired dimensions by a wire saw from a centimeter-size mother sample grown by the vertical Bridgman method~\cite{Li2018}. A large-size sample (\#1) was used for the magnetization measurements and three  samples (\#2, \#3, \#4) with different cross-sections were used for the magnetic torque measurements (See the table~\ref{table-sample} for details). We also studied two bcc Fe samples, cut from the same commercial single crystal, orientated along (110). One sample was cut to a thin-disk shape, with dimensions of $1.3 ~mm$ (diameter) $\times$ $0.12 ~mm$ (thickness). Another sample was cut to a cuboid shape, with dimensions of $0.39 ~mm \times 0.36 ~mm \times 0.6 ~mm$.

All magnetization measurements were performed in a physical property measurement system (Quantum design PPMS-16) with a vibrating sample magnetometer (VSM) option. All magnetic torque measurements were performed in a TeslatronPT (Oxford Instruments) with a homemade rotated probe equipped with the capacitive torque magnetometer. A capacitance bridge (AH-2550A) was used to measure the variation of the capacitance between two metal plates proportional to change in the magnetic torque of the sample. Measurement of the rotation angle was calibrated by a Hall chip.

\begin{table}[htb]
\begin{tabular}{|c|c|c|c|c|}
\hline
 & \#1 & \#2 & \#3 & \#4 \\
\hline
Cross-section & Square & Triangle & Square & Rectangular \\
\hline
$l_x/l_y$ & $\sim$1 & 1.15 & 1.2 & 5.6 \\
\hline
Mass (mg) & 30.0 & 3.12 & 4.4 & 1.15 \\
\hline
Technique & Magnetization & Torque & Torque & Torque \\
\hline
$K_2$ at 5T ($kJ/m^3$) & $\backslash$ & 0.10 & 0.11 & 0.34 \\
\hline
\end{tabular}
\caption{Details of Mn$_3$Sn samples used in this work.}
\label{table-sample}
\end{table}

\begin{figure}
\includegraphics[width=16cm]{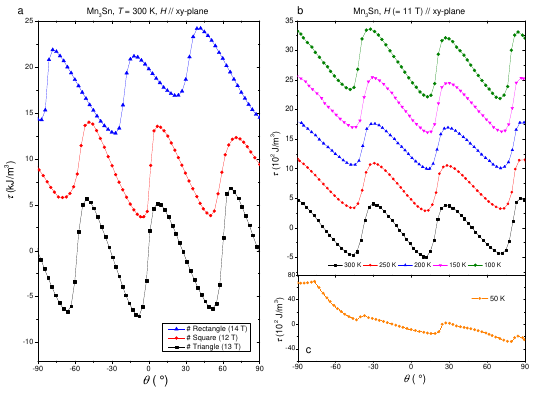}
\caption{Reproducibility of data: The magnetic torque responses in different Mn$_3$Sn samples, with triangle, square, rectangular cross-section respectively (a). Temperature dependent magnetic torque responses in Mn$_3$Sn samples (\# triangle), varying from 300 K to 100 K (b), and 50K (c).}
\label{S1}
\end{figure}

\begin{figure}
\includegraphics[width=16cm]{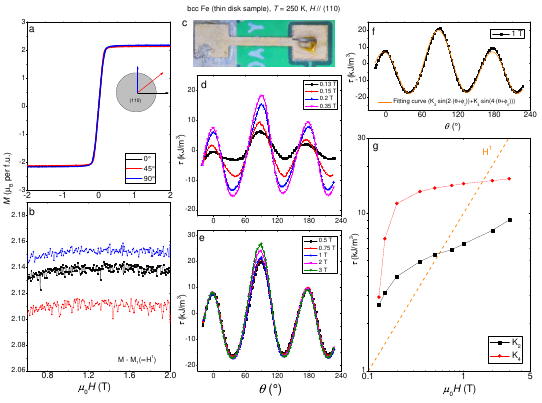}
\caption{Magnetization and the magnetic torque in bcc Fe single crystal (thin disk sample): (a-b) Magnetization. Saturating at 0.2 T (a). No observable quadratic term (b). The demagnetizing factor is approach to 0, in the plane direction of the thin disk. (c-g) The magnetic torque. The photograph of the measurements, the thin disk sample is fixed with a quartz holder(c). The angular torque responses at different field, varying from 0.13 T to 0.35 T (d), 0.5 T to 3 T (e). The fit of the data at 1 T, with the formula ($\tau = K_2 \cdot sin(2 \cdot (\theta + \phi_2)) + K_4 \cdot sin(4 \cdot (\theta + \phi_4))$) (f). Field dependence of $K_2$ and $K_4$ (g).}
\label{S2}
\end{figure}

\begin{figure}
\includegraphics[width=16cm]{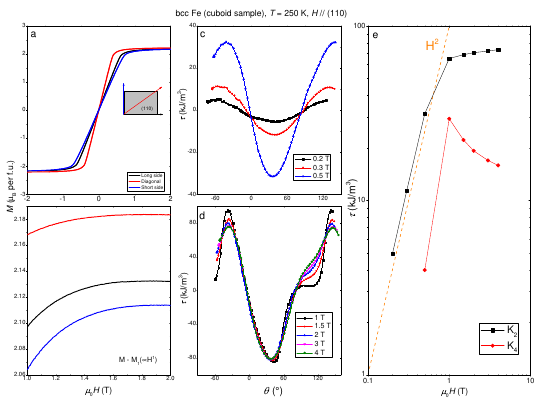}
\caption{Magnetization and the magnetic torque in bcc Fe single crystal (cuboid sample). (a-b) Magnetization. Saturating at 0.5 T (a). No observable quadratic term (b). (c-e) The magnetic torque. The angular torque responses at different field, varying from 0.2 T to 0.5 T (c), 1T to 4 T (d). Field dependence of $K_2$ and $K_4$ (e).}
\label{S3}
\end{figure}


\section{Reproducibility of data} 

As seen in Fig.~\ref{S1}a, we repeated the torque measurements in three Mn$_3$Sn samples with different cross-section. All the samples show the same six-fold oscillation accompanied by a deviation at high field, suggesting the $K_6$ and $K_{12}$ are the intrinsic responses in Mn$_3$Sn. While in the  sample \#4 (with a rectangular cross-section) the two-fold oscillation was more prominent. This indicates that the origin of $K_2$ is linked to the boundary geometry. As argued previously~\cite{Li2018,Li2021}, there is a lag angle between the applied magnetic field and magnetization in Mn$_3$Sn and this lag angle increases with increasing aspect ratio of the sample~\cite{Li2018,Li2021}. This provides an explanation for the larger $K_2$ in the sample with the largest aspect ratio (Fig.~\ref{fig: Magnetic anisotropic energy}e and see the table~\ref{table-sample} for details). 

In the case of the sample with a triangular cross section, we  repeated the torque measurements at different temperatures. The data is presented in Fig.~\ref{S1}b-c. When we compare the high field torque data at different temperature, we find the $K_6$ and $K_{12}$ are robust from 300 K down to 100 K, as seen in Fig.~\ref{S1}b. But the signal collapsed suddenly at 50 K, nearing the transition temperature of the spin glass state~\cite{Nakatsuji2015}, as seen in Fig.~\ref{S1}c. Our data suggests that the angle-dependent magnetic free energy has little temperature dependence and implies that this is a property of the triangular spin texture of the AF state which is destroyed at low temperature. 


\section{Magnetization and magnetic torque in bcc iron} 

In order to compare with Mn$_3$Sn, we measured the magnetization of a disk-shape bcc Fe single crystal at 250 K,  with the field along (110) plane. As seen in Fig.~\ref{S2}a-b. Unlike  Mn$_3$Sn, the dominant component in bcc Fe is the spontaneous magnetization, which saturates at 0.2 T, as seen in Fig.~\ref{S2}a. Focusing on the signal at high field region and having subtracted a linear background no  quadratic magnetization is detectable (Fig.~\ref{S2}b).

We also measured the magnetic torque in the same crystal at 250 K, with the field rotating in (110) plane. Fig.~\ref{S2}c shows the photograph of the measurements. Fig.~\ref{S2}d-e show the angular torque responses in presence of different magnetic fields. A clear four-fold oscillations combined with the two-fold component can be observed. Fig.~\ref{S2}f shows the fit of the data at 1 T. From which we extracted the amplitude of two-fold ($K_2$) and four-fold ($K_4$) components, and plotted them as a function of magnetic field in Fig.~\ref{S2}g. It can be seen that the dominant $K_4(H)$ begin to saturate above 0.2 T, same to the literature~\cite{Graham1958}, but the $K_2(H)$ still maintained a slow growth. 

Comparison of the torque angular oscillations in bcc Fe and in Mn$_3$Sn reveals a fundamental difference. In Fe, the toque (that is the angular dependence of the magnetic free energy) does not change with magnetic field (at least above the coercive field). In Mn$_3$Sn, torque increases strongly with magnetic field ($\propto H^3$). In the first case, the angular dependence of magnetic free energy is caused by single-ion-anisotropy and is field independent. In the second case, the larger the magnetic field, the larger the field-induced twist of the spins and the larger the energy cost of the rotation.


We also repeated the magnetization and magnetic torque of bcc Fe single crystal in another cuboid sample. As seen in Fig.~\ref{S3}, the magnetization responses have the higher saturation field ($H_s$), about 0.5 T, and the in-plane curves are obviously separated, suggesting a larger anisotropy originated from the shape anisotropy. In contrast to the case in Fig.~\ref{S2}, the torque responses in cuboid sample show a obvious two-fold oscillation, and the four-fold component only emerges above 1 T, as seen in Fig.~\ref{S3}. Moreover, the $K_2(H)$ is about ten times larger, imply the shape anisotropy can have a remarkable contribution to the magnetic anisotropy in bcc Fe. However, it doesn't appear in Mn$_3$Sn case, as seen in Fig.~\ref{S1}.

\section{Comparison with previous studies on bcc iron}

Our data is compatible with what was previously reported in the case of bcc Fe. However,  there is a difference of notations. The parameters K$_i$ used in this work are distinct from those traditionally used in the case of cubic ferromagnets. In the latter case, the free energy is written as 
\begin{equation}
  \label{eq:trad}
  E=K_0+K_1^*(\alpha_1^2\alpha_2^2+\alpha_2^2\alpha_3^2+\alpha_3^2\alpha_1^2)+K_2^*(\alpha_1^2\alpha_2^2\alpha_3^2)+...
\end{equation}
Here $\alpha_i$ are direction cosines of the spontaneous magnetization with respect to the cubic axes \cite{Bozorth1936}. How, when ($hkl$) = (110), the torque can be written as $\tau = K_1^* (2 sin2\theta + 3 sin4\theta)/8 + K_2^* (sin2\theta + 4 sin4\theta - 3 sin6\theta)/64$. Therefore, there is a simple relationship between K$_i^*$ and K$_i$: $K_1^* = \frac{8}{5} (4 K_2 - K_4)$ and $K_2^* = \frac{64}{5}(2 K_4 - 3 K_2)$. 

As seen in Fig.~\ref{S2}f,  K$_4$ saturates to about 17 kJ/m$^3$ (1 kJ/m$^3$ = 10$\cdot$ 10$^3$ ergs/cm$^3$). On the other hand the saturated value for K$_2$ can be estimates to be 11 kJ/m$^3$. This would yield a K$_1^*$ and K$_2^*$ of 43.2 and 12.8 kJ/m$^3$. As an see in the table~\ref{table-iron expression}, these values are close to what was reported previously \cite{Bozorth1936, Graham1958}.

\begin{table}[htb]
\begin{tabular}{|c|c|c|c|}
\hline
 Component & Expression & This work value (kJ/m$^3$) & Previously reported (kJ/m$^3$)  \\
\hline
$\tau$ (This work) & $K_2 \cdot sin2\theta + K_4 \cdot sin4\theta $ \cite{Bozorth1936} & $\backslash$ & $\backslash$ \\
\hline
$\tau$ (Literature) & $K_1^* \cdot (2 sin2\theta + 3 sin4\theta)/8 + K_2^* \cdot (sin2\theta + 4 sin4\theta - 3 sin6\theta)/64$ & $\backslash$ & $\backslash$ \\
\hline
K$_2$ & $K_1^* / 4 + K_2^* / 64$ & 11 (estimated)& $\sim$ 12 at 2 T\cite{Graham1958}\\
\hline
K$_4$ & $ 3 K_1^* / 8 + K_2^* / 16$ & 17  & $\sim$ 16 at 2 T \cite{Graham1958}\\
\hline
K$_1^*$ & $\frac{8}{5} (4 K_2 - K_4)$ & 43.2 & 40-52.5 \cite{Bozorth1936} or 48 $\pm$ 1 \cite{Graham1958}\\
\hline
K$_2^*$ & $\frac{64}{5}(2 K_4 - 3 K_2)$ & 12.8 & -17-29 \cite{Bozorth1936} or 0 $\pm$ 5 \cite{Graham1958} \\
\hline
\end{tabular}
\caption{Torque's expressions and values for bcc Fe in (110) plane.}
\label{table-iron expression}
\end{table}

\section{Theory of energy from microscopic spin model}
\label{sec:details-expansion}

Here we extend the small anisotropy expansion of the earlier paper of
Liu and Balents to higher order in the presence of an applied field.
We consider the bulk energy of a uniform state.  The basic unit of the
structure is a set of 6 spins, forming a triangle in one layer and its
counterpart in the neighboring layer.  The spins in the second layer
are assumed to be equal to their counterparts in the first layer.
Then the energy depends upon three so far unspecified spins $\bm{S}_i$
with $i=1,2,3$.  Then we write the energy {\em per unit cell --
  i.e. per 6 spins} as
\begin{equation}
  \label{eq:1}
  E_{u.c.} = \sum_{i} \left[ 4J \bm{S}_i\cdot\bm{S}_{i+1} - 2 K 
  (\bm{S}_i\cdot\bm{\hat{e}}_i)^2 + 4 D \bm{\hat{z}}\cdot \bm{S}_i
  \times \bm{S}_{i+1}-2\mu \bm{H}\cdot \bm{S}_i\right] ,
\end{equation}
where we use the convention $\bm{S}_{i+3}=\bm{S}_i$.  The factors
account for the two copies of the triangles etc.  We take the spins to
be unit vectors so factors of spin length and magnetic moments should
be absorbed into the above parameters.  The parameter $J=J_1+J_2$.

We assume $K \ll D \ll J$ and $\mu H\ll J$, and $D>0$, and obtain an expansion of the
ground state energy.  For $D>0$, the ground state for $K=\mu H=0$ is an
anti-chiral state with spins in the plane.

\subsection{Expansion to fourth order for in-plane fields}
\label{sec:expans-sixth-order}

The most important case is for an in-plane field.  So take
\begin{equation}
  \label{eq:2}
  \bm{H} = H (\cos\theta,\sin\theta,0).
\end{equation}
We can further assume in this case that the spins remain in the plane
even for non-zero $K$ and $H$, hence
\begin{equation}
  \label{eq:3}
  \bm{S}_i = (\cos\phi_i,\sin\phi_i,0).
\end{equation}
We then write the angles as
\begin{equation}
  \label{eq:4}
  \phi_1 = \phi+\eta_1, \qquad \phi_2 = \phi -\frac{2\pi}{3} + \eta_2,
  \qquad \phi_3 = \phi -\frac{4\pi}{3} -\eta_1-\eta_2.
\end{equation}
Here $\phi$ gives the order parameter angle, and $\eta_1,\eta_2$ are
small distortions of the triangle.

To be systematic, we introduce a small parameter $r\ll 1$, and let $K
\rightarrow K r$ and $H \rightarrow H r$, and then expand $\eta_{1,2}$
in a formal series in $r$ and minimize the energy order by order in
$r$.  This is effectively an expansion in $K/J$ and $\mu H/J$, which may
safely be considered small parameters.  There is a priori no need to
assume $D\ll J$ as a second small parameter but since in reality it is
small, it is sometimes convenient to simplify very cumbersome algebraic
expressions, and we will occasionally use it.

This procedure can be carried out at fixed $\phi$.  

So assuming this condition, we can systematically write
\begin{equation}
  \label{eq:6}
  \eta_i = \sum_{n=1}^\infty \eta_{i,n} r^n,
\end{equation}
and expand the full energy order by order in $r$.  We successively
minimize terms beginning at $O(r^2)$ in the energy over $\eta_{i,n}$
which appear in these expressions.  This determines the small
distortions of the triangle at each order and results in a fully determined
expansion of the energy:
\begin{equation}
  \label{eq:7}
  E_{u.c.} = \sum_{n=0}^\infty E_{u.c.}^{(n)},
\end{equation}
where
\begin{align}
  \label{eq:8}
  E_{u.c.}^{(0)}  =  & - 6 J - 6\sqrt{3} D, \\
  E_{u.c.}^{(1)}  = & -3K, \\
  E_{u.c.}^{(2)}  = & -\frac{(\mu H)^2+K^2+ 2 \mu H K \cos(\theta+\phi)}{2 \left(\sqrt{3} D+J\right)}, \\
  E_{u.c.}^{(3)}  = &- \frac{1}{36(J+\sqrt{3}D)^3}\Big[(3J + 7 \sqrt{3}D)K^3\cos(6\phi) +
              6(J+3\sqrt{3}D)\mu H K^2 \cos(5\phi-\theta) \nonumber \\
                & +3
              (J+5\sqrt{3}D)(\mu H)^2K \cos(4\phi-2\theta)+ 4\sqrt{3} D(\mu H)^3 \cos(3\phi-3\theta)\Big].
\end{align}
These expressions are a bit complicated and rather ugly at higher
orders.  Higher order terms are negligible for the effects we discuss here.

The energy should now be minimized over the order parameter angle
$\phi$.  One can see that the third and second order terms have
different angular dependence, hence there is a competition between the
two in determining this angle.

To proceed, we will assume that the second order term is dominant,
being lower order.  This is true {\em unless} $H$ becomes very small,
because the second order term's angular dependence vanishes for
$H=0$.  Comparing the coefficient of $\cos(\phi+\theta)$ from the
second order term and the coefficient of $\cos(6\phi)$ from the
third order term, we see this implies the condition
\begin{equation}
  \label{eq:13}
  \mu H \gg K^2/J.
\end{equation}
When this is true, (and recall we assumed $\mu H,K \ll J$) the second order term is
parametrically larger, so $\phi$ must be close (but not equal!) to the
minimum of this term.  Hence we can write $\phi = -\theta+\psi$, and
we expect $\psi\ll 1$.  Therefore we expand $E_{u.c.}^{(2)}$ to second
order in $\psi$ (it is quadratic around its minimum) and
$E_{u.c.}^{(3)}$ to first order in $\psi$, and minimize over $\psi$.
This leads to the final expression for the energy
\begin{align}
  \label{eq:5}
  E_{u.c.} =   - 6 J - 6\sqrt{3} D - 3K &  -
               \frac{(\mu H +K)^2}{2(J+\sqrt{3}D)} \Big[  1+
               \frac{(3J+7\sqrt{3}D)K + 4\sqrt{3}D \mu H}{18
               (J+\sqrt{3}D)^2 }\cos (6\theta) \nonumber \\
                                                    & +
                                                      \frac{\left((3J+7\sqrt{3}D)K^2+
                                                      2(J+4\sqrt{3}D)\mu H K+2\sqrt{3}D(\mu H)^2\right)^2}{36(J+\sqrt{3}D)^4 \mu H K}\sin^2(6\theta)\Big] .
\end{align}
We see that by including the small shift in the angle ($\psi$), we
generated a 12-fold harmonic (contained in the $\sin^2(6\theta)$
factor.  If we take the high field limit, $\mu H\gg K$ (this is a stronger
condition than Eq.~\eqref{eq:13}) the expression simplifies further to
\begin{align}
  \label{eq:14}
    E_{u.c.} =   - 6 J - 6\sqrt{3} D - 3K  -
               \frac{(\mu H)^2}{2(J+\sqrt{3}D)} \Big[  & 1+
               \frac{2\sqrt{3}D \mu H}{9
               (J+\sqrt{3}D)^2 }\cos (6\theta)  +
                                                      \frac{D^2(\mu H)^3}{3(J+\sqrt{3}D)^4 K}\sin^2(6\theta) \Big].
\end{align}

In solving for the energy, we obtain the angles $\eta_1,\eta_2$ and $\psi$:
\begin{align}
    \eta_1 = & \frac{\mu H+K}{3(J+\sqrt{3}D)} \Big[\sin 2\theta - \frac{\mu H+2K}{6\mu H K(J+\sqrt{3}D)^2} \left(2\sqrt{3}D(\mu H)^2+2(J+4\sqrt{3}D)\mu HK+ (3J+7\sqrt{3}D)K^2\right)\cos2\theta \sin6\theta\Big], \nonumber \\
    \eta_2 = & \frac{\mu H+K}{3(J+\sqrt{3}D)} \Big[\cos(2\theta+\frac{\pi}{6}) \nonumber \\
    &+ \frac{\mu H+2K}{6BK(J+\sqrt{3}D)^2} \left(2\sqrt{3}D (\mu H)^2+2(J+4\sqrt{3}D)\mu HK+ (3J+7\sqrt{3}D)K^2\right)\cos(2\theta-\frac{\pi}{3})\sin6\theta\Big],\nonumber \\
    \psi = & \frac{B+K}{6\mu HK(J+\sqrt{3}D)^2} \left(2\sqrt{3}D(\mu H)^2+2(J+4\sqrt{3}D)\mu HK+ (3J+7\sqrt{3}D)K^2\right)\sin 6\theta.
\end{align}

Despite the complexity, here are a few features:
\begin{itemize}
\item A $H^2$ term appears first at order $n=2$,
and has no angular dependence.  Because terms successively decrease
with increasing $n$, this is the dominant contribution to the $H^2$
term.  It represents  term in the magnetization linear in field,
i.e. just a susceptibility.  The $D$ in the denominator is
subdominant, and so the susceptibility is just order $1/J$. 
\item The $H^3$ term appears first at order $n=3$.  Examining this
expression we see that this term however is linear in $D$ and vanishes
for $D=0$.  The third order contribution has a pure $\cos 6\theta$ dependence.
\item The $\cos 12\theta$ harmonic is formally of fourth order
  (e.g. in Eq.~\eqref{eq:14} its coefficient is order $H^5/K$).  One
  can check that no additional 12-fold contribution arises if we
  include the $E_{u.c.}^{(4)}$ correction. 
\end{itemize}

Many of these features may match experimental observations.  A few notable differences from theory are:
\begin{itemize}
\item There are no 2-fold harmonics.  They can appear only
  from effects that break the six-fold structural symmetry of the
  solid, e.g. finite size shape effects, or strain due to
  magnetostriction in the ordered state.
\item The energy is analytic in $H$.  This means that the fractional power appearing experimentally in the two-fold anisotropy must arise from additional physics not included here, which is hardly surprising since the two-fold anisotropy itself is not present in this model.
  \item The $H^3$ term has a pure $\cos(6\theta)$ dependence on the angle, i.e. there is no angle-independent part.  A small angle-independent part does arise at higher orders in the expansion but it is negligible.  This means that the angle-independent $H^3$ term seen in experiment must arise from effects beyond the simple model studied here.  
\end{itemize}

\subsection{Arbitrary spin angle}
\label{sec:arbitrary-spin-angle}

It is too cumbersome to carry out the full expansion to 6th order for
applied fields in arbitrary directions, but we can do so for the first
few orders.  In this case, we can repeat similar manipulations but not
assuming in-plane spins.  Suppressing the details, we obtain the
following energy {\em to second order in $H,K$}:
\begin{align}
  \label{eq:9}
  E_{u.c.} = & -6J - 6\sqrt{3}D - 3K  - \frac{1}{2J}\left(1 -
  \frac{\sqrt{3}D}{J}\right) (K +
  \mu H_{xy})^2 \nonumber \\
  &- \frac{1}{2J}\left(1 - \frac{D}{\sqrt{3}J}\right) (\mu H_z)^2+ O(r^3).
\end{align}
From this we can see that the linear in field magnetization is almost
isotropic, i.e. the fractional difference in differential susceptibility in the
two directions is order $D/J$.

\end{document}